%% file: WASP-72_WASP-100_WASP-109_RM_V3.tex
\documentclass[iop, apj]{emulateapj}
\usepackage[]{hyperref}
\usepackage{textcomp}
\usepackage{placeins}
\usepackage{multirow}
\usepackage{graphicx}
\usepackage{amsmath}
\usepackage{amssymb,latexsym}
\usepackage{threeparttable}
\usepackage{longtable}
\usepackage{mhchem}
\input{aas_macros.tex}

\usepackage{natbib}
\usepackage{adjustbox}

\hypersetup{
     colorlinks   = true,
     citecolor     = blue
} 

\newcommand{\mos}{\,m\,s$^{-1}$}
\newcommand{\kms}{\,km\,s$^{-1}$}

\newcommand\msini{\ifmmode{{\mathrm M} \sin i}\else${{\mathrm M} \sin i}$\fi}

\slugcomment{Accepted for publication in The Astronomical Journal}

\shorttitle{Spin-Orbit measurements of Three Transiting Hot: WASP-72\MakeLowercase{b}, WASP-100\MakeLowercase{b}, \& WASP-109\MakeLowercase{b}}
\shortauthors{Addison et al.}

\begin{document}

\title{Stellar Obliquities \& Planetary Alignments (SOPA) I. Spin-Orbit measurements of Three Transiting Hot Jupiters: WASP-72\MakeLowercase{b}, WASP-100\MakeLowercase{b}, \& WASP-109\MakeLowercase{b}\altaffilmark{$^{\dagger}$}}

\author{B. C. Addison\altaffilmark{1,2}, S. Wang\altaffilmark{3,$^{\star}$}, M. C. Johnson\altaffilmark{4}, C. G. Tinney\altaffilmark{5,6}, D. J. Wright\altaffilmark{2}, D. Bayliss\altaffilmark{7}}

\email{baddison2005@gmail.com}

\altaffiltext{1}{Mississippi State University, Department of Physics \& Astronomy, Hilbun Hall, Starkville, MS 39762}
\altaffiltext{2}{University of Southern Queensland, Centre for Astrophysics, Toowoomba, Queensland, Australia, 4350}
\altaffiltext{3}{Department of Astronomy, Yale University, New Haven, CT 06511}
\altaffiltext{4}{Department of Astronomy, The Ohio State University, 140 West 18th Ave., Columbus, OH 43210 USA}
\altaffiltext{5}{Exoplanetary Science Group, School of Physics, University of New South Wales, Sydney, NSW 2052, Australia}
\altaffiltext{6}{Australian Centre of Astrobiology, University of New South Wales, Sydney, NSW 2052, Australia}
\altaffiltext{7}{Department of Physics, University of Warwick, Coventry, UK}
\altaffiltext{$\dagger$}{Based on observations obtained at the Anglo-Australian Telescope, Siding Spring, Australia.}
\altaffiltext{$^{\star}$}{\textit{51 Pegasi b} Fellow}

\date{\today}

\begin{abstract}
We report measurements of the sky-projected spin--orbit angles for three transiting hot Jupiters: two of which are in nearly polar orbits, WASP-100b and WASP-109b, and a third in a low obliquity orbit, WASP-72b. We obtained these measurements by observing the Rossiter--McLaughlin effect over the course of the transits from high resolution spectroscopic observations made with the CYCLOPS2 optical fiber bundle system feeding the UCLES spectrograph on the Anglo-Australian Telescope. The resulting sky-projected spin--orbit angles are $\lambda = {-7^{\circ}}^{+11^{\circ}}_{-12^{\circ}}$, $\lambda = {79^{\circ}}^{+19^{\circ}}_{-10^{\circ}}$, and $\lambda = {99^{\circ}}^{+10^{\circ}}_{-9^{\circ}}$ for WASP-72b, WASP-100b, and WASP-109b, respectively. These results suggests that WASP-100b and WASP-109b are on highly inclined orbits tilted nearly $90^{\circ}$ from their host star's equator while the orbit of WASP-72b appears to be well-aligned. WASP-72b is a typical hot Jupiter orbiting a mid-late F star (F7 with $T_\mathrm{{eff}}=6250\pm120$\,K). WASP-100b and WASP-109b are highly irradiated bloated hot Jupiters orbiting hot early-mid F stars (F2 with $T_\mathrm{{eff}}=6900\pm120$\,K and F4 with $T_\mathrm{{eff}}=6520\pm140$\,K), making them consistent with the trends observed for the majority of stars hosting planets on high-obliquity orbits.


\end{abstract}


\section{INTRODUCTION}
\setcounter{footnote}{3}

Despite decades of inquiry, the origin of hot Jupiters remains unclear \citep{2017AJ....154...93S}. The standard paradigm holds that these behemoths were not born \textit{in situ} (for an opposing view, however, see \citealt{2016ApJ...829..114B}), but rather that they formed beyond the protostellar ice line where raw materials are plentiful \citep{2000Icar..143....2B}. They then migrated inward via disk-migration mechanisms \citep{1996Natur.380..606L}, or dynamical-migration mechanisms, including: planet-planet scattering \citep{2008ApJ...686..621F,2008ApJ...678..498N}, Lidov-Kozai cycling with tidal friction \citep{2003ApJ...589..605W,2007ApJ...669.1298F,2011Natur.473..187N}, and secular chaos \citep{2011ApJ...735..109W}. The dominant mechanism of migration, however, remains controversial \citep{2016Natur.534..662D}.

The successful migration scenario has to explain at least two observed properties of hot Jupiters:

First, hot Jupiters are frequently observed to have orbital planes that are misaligned with the equators of their host stars (as reviewed by \citealt{2015ARA&A..53..409W}). This is particularly true for stars hotter than the Kraft break \cite{Winn:2010}, at $T_{\mathrm{eff}}\sim6250$ K \citep{Kraft:1967}. Dynamical migration violently delivers giant planets to their current orbits, and can naturally leave systems misaligned. In this framework, the spin-orbit misalignments should be confined to hot Jupiters. It is still plausible that hot Jupiters formed via quiescent migration, and spin-orbit misalignments might alternatively be excited via independent mechanisms that are unrelated to planet migration. These include chaotic star formation \citep{2010MNRAS.401.1505B,2011MNRAS.417.1817T,2015MNRAS.450.3306F}, angular momentum transport within a host star by internal gravity waves \citep[IGW, see, ][]{2012ApJ...758L...6R}, magnetic torques from host stars \citep{2011MNRAS.412.2790L}, and gravitational torques from distant companions \citep{1991Icar...89...85T,2011A&A...533A...7B,2014Sci...345.1317S}. In these scenarios, the spin-orbit misalignments should occur not only in hot Jupiter systems, but also in a broader class of planetary systems, including, crucially, multi-planet systems that have never experienced chaotic migration. 

Spin-orbit misalignments are usually determined by measuring the Rossiter-McLaughlin effect \citep{1924ApJ....60...15R,1924ApJ....60...22M}, a time-variable anomaly in the stellar spectral-line profiles and hence radial velocity during the transit \citep{2000A&A...359L..13Q}. It is much more easily measured when transits are frequent and deep. Therefore, as a practical consequence, while Rossiter-Mclaughlin observations of multi-planet systems play a critical role in understanding planetary formation history, they are hard to make. They usually involve fainter stars, smaller transit depths, and/or less frequent transits, and as yet, very few high quality measurements exist (Kepler-89 d, \citealt{2012ApJ...759L..36H} and \citealt{2013ApJ...771...11A}; Kepler-25 c, \citealt{2013ApJ...771...11A} and \citealt{2014PASJ...66...94B}; WASP-47 b, \citealt{2015ApJ...812L..11S}； Kepler-9b, \citealt{2018AJ....155...70W}). Hence why the majority of Rossiter-Mclaughlin observations are of hot Jupiters.

The second notable property is that hot Jupiters tend to be alone. Although many hot Jupiters detected with Kepler \citep{2010Sci...327..977B} do not appear to have additional close-in transiting planets \citep{2012ApJ...756..186S,2015MNRAS.454.4159H}, the possible presence of such planets in hot Jupiter systems discovered by ground-based photometric surveys (e.g. SuperWASP, \citealt{2006PASP..118.1407P}; HAT, \citealt{2004PASP..116..266B}; KELT, \citealt{2007PASP..119..923P}; CSTAR, \citealt{2014ApJS..211...26W}), which constitute the major fraction (about two thirds) of all currently known hot Jupiters, has not been ruled out. Neptune-sized planets transiting Sun-like stars cause drops in stellar brightness of $\sim 0.12\%$, which remain somewhat beyond the capabilities of existing ground-based transit surveys. Leading research groups are now typically achieving photometric errors of $\sim$0.4\% with wide-field photometric telescopes. WASP-47b is a typical hot Jupiter that was originally detected with SuperWASP \citep{2012MNRAS.426..739H}. Two additional transiting short-period super-Earths (planets several times Earth's mass) in the system did not show up until follow-up observations were obtained from the \textit{Kepler} spacecraft during its K2 mission \citep{2015ApJ...812L..18B}.

NASA's upcoming TESS mission \citep{2014SPIE.9143E..20R} will perform high-precision photometric follow-up for the majority of known transiting hot Jupiters, and it will provide decisive constraints on the occurrence rate of WASP-47-like systems (that is the occurrence rate of the systems harboring both hot Jupiters and additional close-in planets). We have initiated the Stellar Obliquities \& Planetary Alignments (SOPA) project to characterize the spin-orbit angle distribution for the same sample of systems, the sample of hot Jupiters detected with the ground-based transit surveys but without the Rossiter-McLaughlin measurements. More spin-orbit angle determinations for hot Jupiter systems were originally considered to be gradually losing its cachet. Together with TESS, however, it will for the first time link hot Jupiters' two most notable observable properties, and answer the critical question: what are the dominate mechanism(s) driving the formation, migration, and spin-orbit misalignment of hot Jupiters?

Here we present measurements of the spin-orbit misalignments of three hot Jupiters: WASP-72b (\citealt{2013A&A...552A..82G}), WASP-100b (\citealt{2014MNRAS.440.1982H}), and WASP-109b (\citealt{2014arXiv1410.3449A}). The latter two of these orbit stars above the Kraft break, while WASP-72 is located at the Kraft break.

\section{OBSERVATIONS}

We carried out the spectroscopic observations of WASP-72b, WASP-100b, and WASP-109b using the CYCLOPS2 fiber feed with the UCLES spectrograph on the Anglo-Australian Telescope (AAT). CYCLOPS2 is a Cassegrain fiber-based integral field unit with an equivalent on the sky diameter aperture of $\sim2.5^{\texttt{"}}$, reformated into a pseudo-slit of width $0.6^{\texttt{"}}$ at the entrance of the UCLES spectrograph. It delivers a spectral resolution of $R=70,000$ in the wavelength range of $4550-7350$\AA{} across 19 echelle orders with readout times of 175\,s. The instrumental set up and observing strategy for the transit observations closely followed that presented in our previous Rossiter--McLaughlin publications \citep[i.e., WASP-103b, WASP-87b, \& WASP-66b;][]{2016ApJ...823...29A}. We used a thorium--argon calibration lamp (ThAr) to illuminate all on-sky fibers, and a thorium--uranium--xenon lamp (ThUXe) to illuminate the simultaneous calibration fiber for calibrating the observations. The radial velocity measurements are listed in Tables~\ref{table:spec_WASP-72}, \ref{table:spec_WASP-100}, \& \ref{table:spec_WASP-109}.


\subsection{Spectroscopic Observations of WASP-72b}

To measure the Rossiter-McLaughlin effect of WASP-72b, we obtained time-series spectroscopic observations of the transit on 2014 October 01. Observations began at 13:41UT ($\sim\!60$\,minutes before ingress) and were completed at 18:47UT ($\sim\!15$\,minutes after egress). A total of 18 spectra with an exposure time of 960\,s were obtained on that night (12 during the $\sim\!4$\,hr transit) in average observing conditions for Siding Spring Observatory with seeing varying between $1.1^{\texttt{"}}$ and $1.4^{\texttt{"}}$ under clear skies. The airmass at which WASP-72b was observed at varied between of 1.3 for the first exposure, 1.1 near mid-transit, and 1.3 for the last observation. 

\begin{table}[b]
\centering
\begin{minipage}{\columnwidth}
\caption{Radial velocity observations of WASP-72}
\centering
\resizebox{\textwidth}{!}{%
\begin{tabular}{c c c}
\hline\hline \\ [-2.0ex]
Time [BJD] & Radial velocity [m/s] & Uncertainty [m/s] \\ [0.5ex]
\hline \\ [-2.0ex]
2457297.07965  &  37  &   14   \\
2457297.09214  &  69  &   11   \\
2457297.10463  &  39  &   12   \\
2457297.11712  &  39  &   11   \\
2457297.12961  &  35  &   9   \\
2457297.14211  &  45  &   17   \\
2457297.15460  &  65  &   9   \\
2457297.16709  &  21  &   11   \\
2457297.17958  &  20  &   11   \\
2457297.19207  &  0  &   15   \\
2457297.20456  &  -2  &  13   \\
2457297.21705  &  -13  &  12   \\
2457297.22954  &  -45  &  14   \\
2457297.24203  &  -30  &  9   \\
2457297.25453  &  -37  &  15   \\
2457297.26702  &  -37  &  12   \\
2457297.27951  &  -29  &  14   \\
2457297.29200  &  -70  &  17 \\
\hline \\ [-2.5ex]
\end{tabular}}
\label{table:spec_WASP-72}
\end{minipage}
\end{table}

\subsection{Spectroscopic Observations of WASP-100b}

We obtained spectroscopic observations of the transit of WASP-100b on the night of 2015 October 02, starting 50\,minutes before ingress and finishing 74\,minutes after egress. A total of 18 spectra with an exposure time of 1000\,s were obtained on that night (including 11 during the $\sim\!4$\,hr transit) with clear skies and seeing varying between $0.9^{\texttt{"}}$ and $1.2^{\texttt{"}}$. WASP-100 was observed at an airmass of 2.0 for the first exposure, 1.40 near mid-transit, and 1.2 at the end of the observations.

\begin{table}[b]
\centering
\begin{minipage}{\columnwidth}
\caption{Radial velocity observations of WASP-100}
\centering
\resizebox{\textwidth}{!}{%
\begin{tabular}{c c c}
\hline\hline \\ [-2.0ex]
Time [BJD] & Radial velocity [m/s] & Uncertainty [m/s] \\ [0.5ex]
\hline \\ [-2.0ex]
2457298.02098  &  90  &   29   \\
2457298.03463  &  41  &   22   \\
2457298.04828  &  -20  &  30   \\
2457298.06192  &  -26  &  19   \\
2457298.07557  &  -102 &  27   \\
2457298.08921  &  -103 &  25  \\
2457298.10286  &  -94  &  19   \\
2457298.11650  &  -107 &  15   \\
2457298.13014  &  -96  &  20   \\
2457298.14381  &  -138 &  19   \\
2457298.15745  &  -96  &  30   \\
2457298.17110  &  -123 &  21   \\
2457298.18474  &  -116 &  21  \\
2457298.19840  &  -31  &  24   \\
2457298.21204  &  -60  &  27   \\
2457298.22569  &  -67  &  22   \\
2457298.23934  &  -96  &  28  \\
2457298.25298  &  -49  &  22 \\
\hline \\ [-2.5ex]
\end{tabular}}
\label{table:spec_WASP-100}
\end{minipage}
\end{table}

\subsection{Spectroscopic Observations of WASP-109b}

We observed the transit of WASP-109b on the night of 2015 May 08, starting $\sim\!50$\,minutes before ingress and finishing $\sim\!35$\,minutes after egress. A total of 16 spectra with an exposure time of 900\,s were obtained on that night (10 during the $\sim\!3$\,hr transit) under clear skies but with poor seeing conditions (the seeing varied between $\sim\!1.9^{\texttt{"}}$ to $\sim\!2.8^{\texttt{"}}$). WASP-109 was at an airmass of 1.15 for the first exposure, 1.05 near mid-transit, and 1.3 at the end of the observations.

\begin{table}[b]
\centering
\begin{minipage}{\columnwidth}
\caption{Radial velocity observations of WASP-109}
\centering
\resizebox{\textwidth}{!}{%
\begin{tabular}{c c c}
\hline\hline \\ [-2.0ex]
Time [BJD] & Radial velocity [m/s] & Uncertainty [m/s] \\ [0.5ex]
\hline \\ [-2.0ex]
2457151.03710  &  6    &  52   \\
2457151.04894  &  149  &  81   \\
2457151.06079  &  -115 &  81   \\
2457151.07263  &  -6   &  83   \\
2457151.08448  &  -234 &  95   \\
2457151.09632  &  -462 &  51  \\
2457151.10817  &  22   &  109   \\
2457151.12001  &  -292 &  72   \\
2457151.13186  &  -74  &  111   \\
2457151.14370  &  -639 &  70   \\
2457151.15555  &  -356 &  167   \\
2457151.16739  &  -221 &  108   \\
2457151.17924  &  -155 &  110  \\
2457151.19108  &  -41  &  124   \\
2457151.20295  &  -107 &  111   \\
2457151.21480  &  -267 &  180 \\
\hline \\ [-2.5ex]
\end{tabular}}
\label{table:spec_WASP-109}
\end{minipage}
\end{table}

\section{Rossiter--McLaughlin Analysis} \label{sec:RM_analysis}

To determine the best-fit $\lambda$ (the sky-projected angle between the planetary orbit and their host star's spin axis) values for WASP-72, WASP-100, and WASP-109 from spectroscopic observations of the Rossiter-McLaughlin effect, we used the Exoplanetary Orbital Simulation and Analysis Model \citep[ExOSAM; see][]{2013ApJ...774L...9A,2014ApJ...792..112A,2016ApJ...823...29A}. For the analysis of these three systems, we ran 10 independent Metropolis-Hastings Markov Chain Monte Carlo \citep[MCMC, procedure largely follows from][]{2007MNRAS.380.1230C} walkers for 50,000 accepted iterations to derive accurate posterior probability distributions of $\lambda$ and $v\sin i_{\star}$ and to optimize their fit to the radial velocity data. The optimal solutions for $\lambda$ and $v\sin i_{\star}$, as well as their $1\sigma$ uncertainties, are calculated from the mean and the standard deviation of all the accepted MCMC iterations, respectively.

Tables 4--6 lists the prior values, the $1\sigma$ uncertainties, and the prior type of each parameter used in the ExOSAM model for all three systems. The results of the MCMC analysis and the best-fit values for $\lambda$ and $v\sin i_{\star}$ are also given in Table 4--6.

For the three systems studied here, we fixed the orbital eccentricity ($e$) to 0, the adopted solution in \citet{2013A&A...552A..82G}, \citet{2014MNRAS.440.1982H}, and \citet{2014arXiv1410.3449A}, respectively. We accounted for the uncertainties on $R_{\star}$, $R_{P}$ and the length of the transit by imposing Gaussian priors on the planet-to-star radius ratio ($R_{P}/R_{\star}$) and the ratio between the orbital semi-major axis and radius of the star ($a//R_{\star}$). Gaussian priors were imposed on the quadratic limb darkening coefficients ($q_{1}$) and ($q_{2}$) based on interpolated values from look-up tables in \citet{2011A&A...529A..75C}. 

We incorporated the uncertainties on the mid-transit epoch ($T_{0}$), the orbital period ($P$), impact parameter ($b$), and the stellar velocity semi amplitude ($K$) into our model using Gaussian priors from the literature. Gaussian priors were set on the stellar macro-turbulence ($v_\mathrm{mac}$) parameter for WASP-72 and WASP-109 from \citet{2013A&A...552A..82G} and \citet{2014arXiv1410.3449A}, respectively. \citet{2014MNRAS.440.1982H} does not provide a value for $v_\mathrm{mac}$ for WASP-100, therefore, we use a reasonable range for our uniform prior between the interval of $0$\kms\, to $10$\kms. The radial velocity offsets ($V_{d}$) between the data we obtained on the AAT and the RVs published in the literature for WASP-72, WASP-100, and WASP-109 were determined using a uniform prior on reasonable intervals as given in Tables~\ref{table:WASP-72_Parameters}, \ref{table:WASP-100_Parameters}, \& \ref{table:WASP-109_Parameters}.  

For $\lambda$, we used uniform priors on the intervals given in Tables~\ref{table:WASP-72_Parameters}, \ref{table:WASP-100_Parameters}, \& \ref{table:WASP-109_Parameters}. These intervals were selected based on the visual inspections of the Rossiter-McLaughlin Doppler anomaly from the time series radial velocities covering each of the transit events. We performed the MCMC analysis using three different priors on ${v\sin i_{\star}}$ based on the values given in \citet{2013A&A...552A..82G}, \citet{2014MNRAS.440.1982H}, and \citet{2014arXiv1410.3449A} for WASP-72, WASP-100, and WASP-109, respectively. The priors used are a normal prior (the reported $v\sin i_{\star}$ and associated $1\sigma$ uncertainty), a weak prior (the reported $v\sin i_{\star}$ and a $3\sigma$ uncertainty), and a uniform prior. Our preferred solution for all three systems is the one using the weak prior on $v\sin i_{\star}$. The weak $v\sin i_{\star}$ prior allows the MCMC to sufficiently explore the parameter space and fit for $\lambda$ and $v\sin i_{\star}$ while incorporating prior information on $v\sin i_{\star}$ as reported in the discovery publications that they obtained from high S/N, high-resolution out-of-transit spectra and constraining the MCMC to sensible $v\sin i_{\star}$ regions.

\subsection{WASP-72 Results} \label{sec:wasp-72_results}

We determined the best-fit projected spin-orbit angle for WASP-72 using the normal $v\sin i_{\star}$ prior of $v\sin i_{\star}=6.0 \pm 0.7$\,\kms\ as $\lambda = {-6^{\circ}}^{+10^{\circ}}_{-12^{\circ}}$. Our preferred solution using the weak $v\sin i_{\star}$ prior of $v\sin i_{\star}=6.0 \pm 2.1$\,\kms\ results in $\lambda = {-7^{\circ}}^{+11^{\circ}}_{-12^{\circ}}$. The best-fit projected spin-orbit angle using a uniform prior on $v\sin i_{\star}$ of $U[1.0 - 12.0]$\,\kms\ is $\lambda = {-7^{\circ}}^{+11^{\circ}}_{-13^{\circ}}$. It should be noted that the inclination of the stellar spin-axis cannot be determined with existing data, therefore, the true spin-orbit angle ($\psi$) is not known \citep[e.g., see][]{2009ApJ...696.1230F}. The results for the stellar rotational velocity are $v\sin i_{\star} = 5.8 \pm 0.7$\,\kms, $v\sin i_{\star} = 5.0^{+1.4}_{-1.2}$\,\kms, and $v\sin i_{\star} = 4.7^{+1.7}_{-1.3}$\,\kms, respectively, for the normal, weak, and uniform prior on $v\sin i_{\star}$. The spin-orbit angle solution does not appear to be affected by the type of $v\sin i_{\star}$ prior used due to the planet's high impact parameter of $b=0.59^{+0.10}_{-0.18}$. $\lambda$ and $v\sin i_{\star}$ are usually less strongly correlated with one another if the impact parameter is high (a more grazing transit), therefore allowing a more precise determination of $\lambda$ \citep{2017arXiv170906376T}.

\begin{figure}
	\centering
	\includegraphics[width=1.0\linewidth]{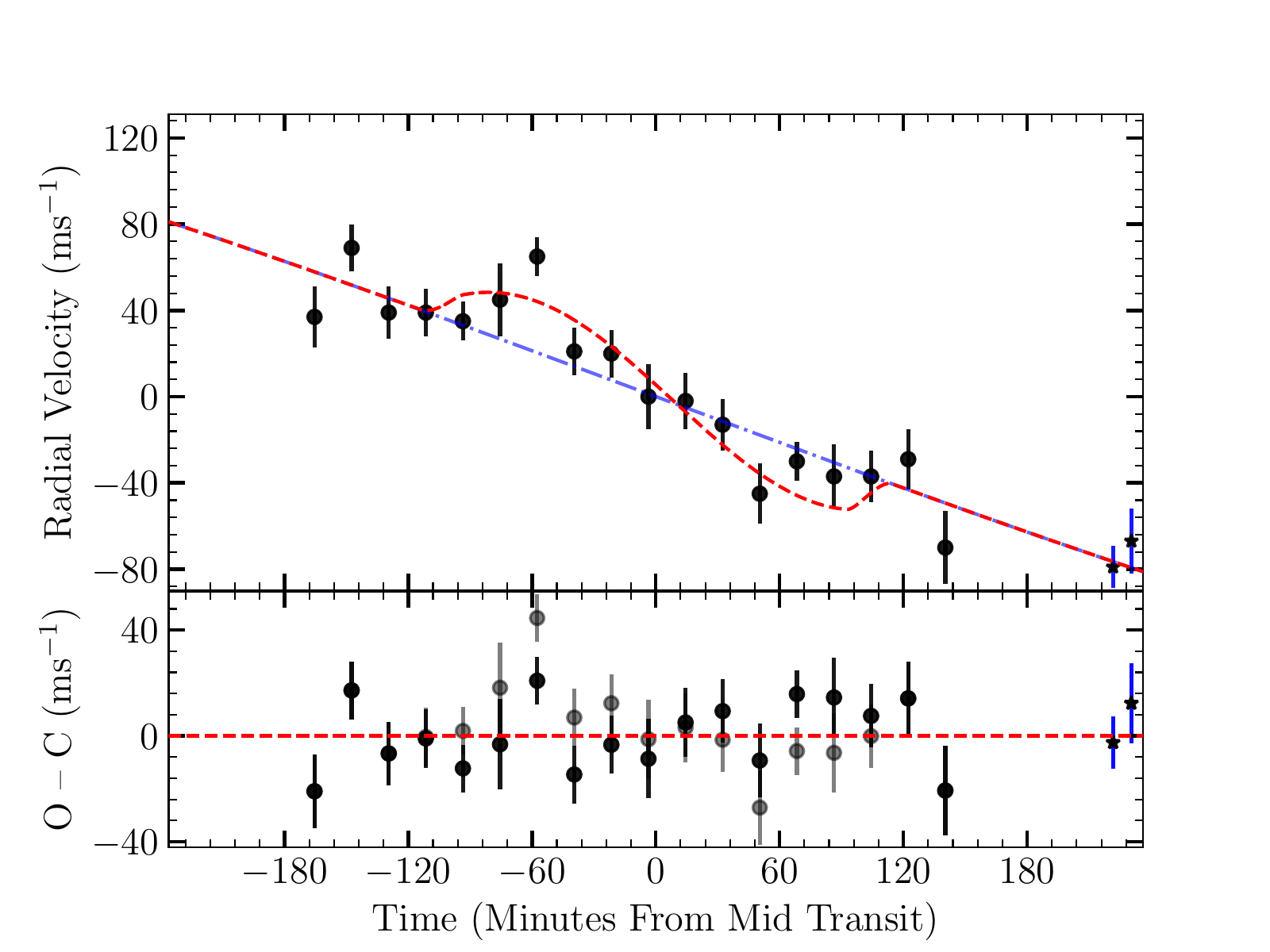}
	\caption{Spectroscopic radial velocities of the WASP-72 transit. Velocities from just before, during, and after the transit are plotted as a function of time (minutes from mid-transit at 2457297.194549\,HJD) along with the best fitting Rossiter-McLaughlin model (using the weak $3\sigma$ ${v\sin i_{\star}}$ prior, our preferred solution), Doppler model with no Rossister-McLaughlin effect, and corresponding residuals. The filled black circles with red error bars are radial velocities obtained in this work on 2015 October 1, the black circles in the residuals plot are from the best fit Rossiter-McLaughlin model, and the gray circles are the residuals from the Doppler model with no Rossister-McLaughlin effect. The two black circles with \textborn\, and with blue error bars are previously published velocities by \citet{2013A&A...552A..82G} using their quoted uncertainties. The velocity offset for the data set presented here was determined from the \citet{2013A&A...552A..82G} out-of-transit radial velocities.}
	\label{fig:WASP-72_RM}
\end{figure}

Our results suggest that the orbit of WASP-72b is aligned to the spin-axis of its host star, assuming the stellar spin-axis is nearly aligned with the sky plane. Figure~\ref{fig:WASP-72_RM} shows the time-series radial velocities during the transit of WASP-72b, the best fit Rossiter-Mclaughlin effect solution, and the residuals to both the best fit Rossiter-McLaughlin model (the black points) and a Doppler solution assuming no Rossiter-McLaughlin effect (the gray points). The Rossiter-McLaughlin effect signal is difficult to discern in the data though a pro-grade solution is evident (seen as a nearly symmetrical velocity anomaly). Therefore, one might wonder how our solution for the spin-orbit angle has such a small uncertainty of only $\Delta \lambda \sim \pm 12^{\circ}$.

\citet{2013ApJ...771...11A} analyzed a similarly low-amplitude Rossiter-McLaughlin effect signal for the Kepler-25 system provides a good explanation for the precise spin-orbit angle solution of WASP-72. As with the Kepler-25 system \citet{2013ApJ...771...11A} analyzed, we have a great deal of prior knowledge of all the system parameters relevant for the Rossiter-McLaughlin effect, with the exception of $\lambda$. This allows us to predict accurately the expected characteristics of the Rossiter-McLaughlin anomaly as a function of $\lambda$. To first order, the amplitude of the Doppler anomaly is proportional to the surface area covered by the transiting planet and the projected rotational speed of the host star. The amplitude of the Rossiter-McLaughlin effect is also strongly dependent on $\lambda$ itself. The amplitude of the Rossiter-McLaughlin signal is larger for polar orbits ($\lambda = \pm90^{\circ}$) than it is for $\lambda$ near 0 degrees or 180 degrees. Additionally, there is a hint of a pro-grade signal in the radial velocity data. Given these factors, the low projected obliquity is strongly favored with a relatively small uncertainty.
 
We also examined in further detail whether the Rossiter-McLaughlin effect signal is actually detected or if a Doppler solution assuming no Rossiter-McLaughlin effect is preferred from the data. To do this, we calculated the Bayesian information Criterion \citep[BIC,][]{1978AnSta...6..461S} and compared the BIC between the two models, finding $\Delta \mathrm{BIC}=9.4$. This gives us strong evidence \citep{doi:10.1080/01621459.1995.10476572} against the null hypothesis (no Rossiter-McLaughlin effect detected) in favor of the Rossiter-McLaughlin model.



\begin{figure}
	\centering
	\includegraphics[width=1.0\linewidth]{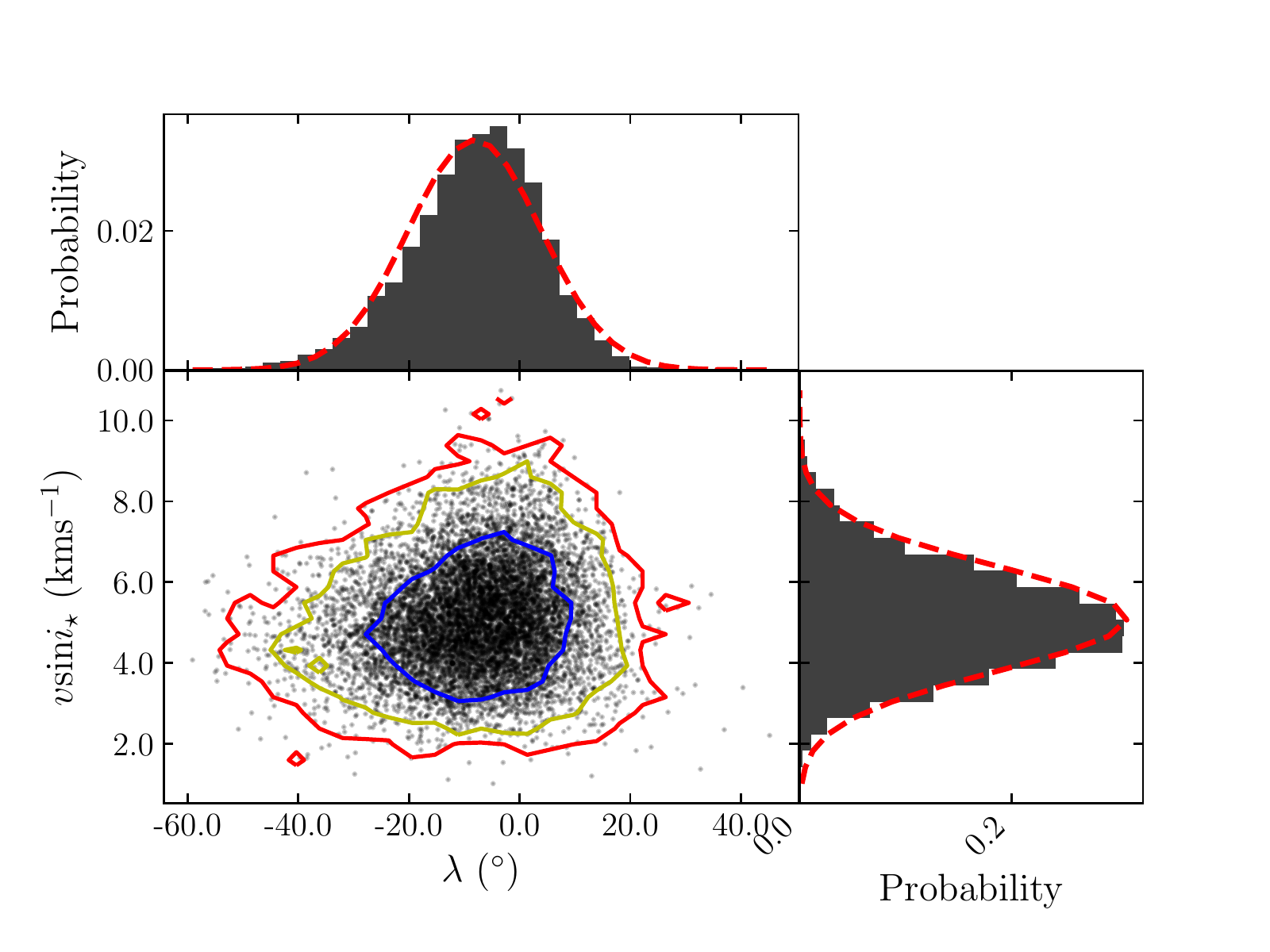}
	\caption{Posterior probability distribution of $\lambda$ and $v\sin i_{\star}$ from the MCMC simulation of WASP-72. The contours show the 1, 2, and 3 $\sigma$ confidence regions (in blue, yellow, and red, respectively). We have marginalized over $\lambda$ and $v\sin i_{\star}$ and have fit them with Gaussians (in red). This plot indicates that the distribution is mostly Gaussian suggesting only a weak correlation between $\lambda$ and $v\sin i_{\star}$.}
	\label{fig:wasp-72_distro}
\end{figure}

\begin{figure}
	\centering
	\includegraphics[width=1.0\linewidth]{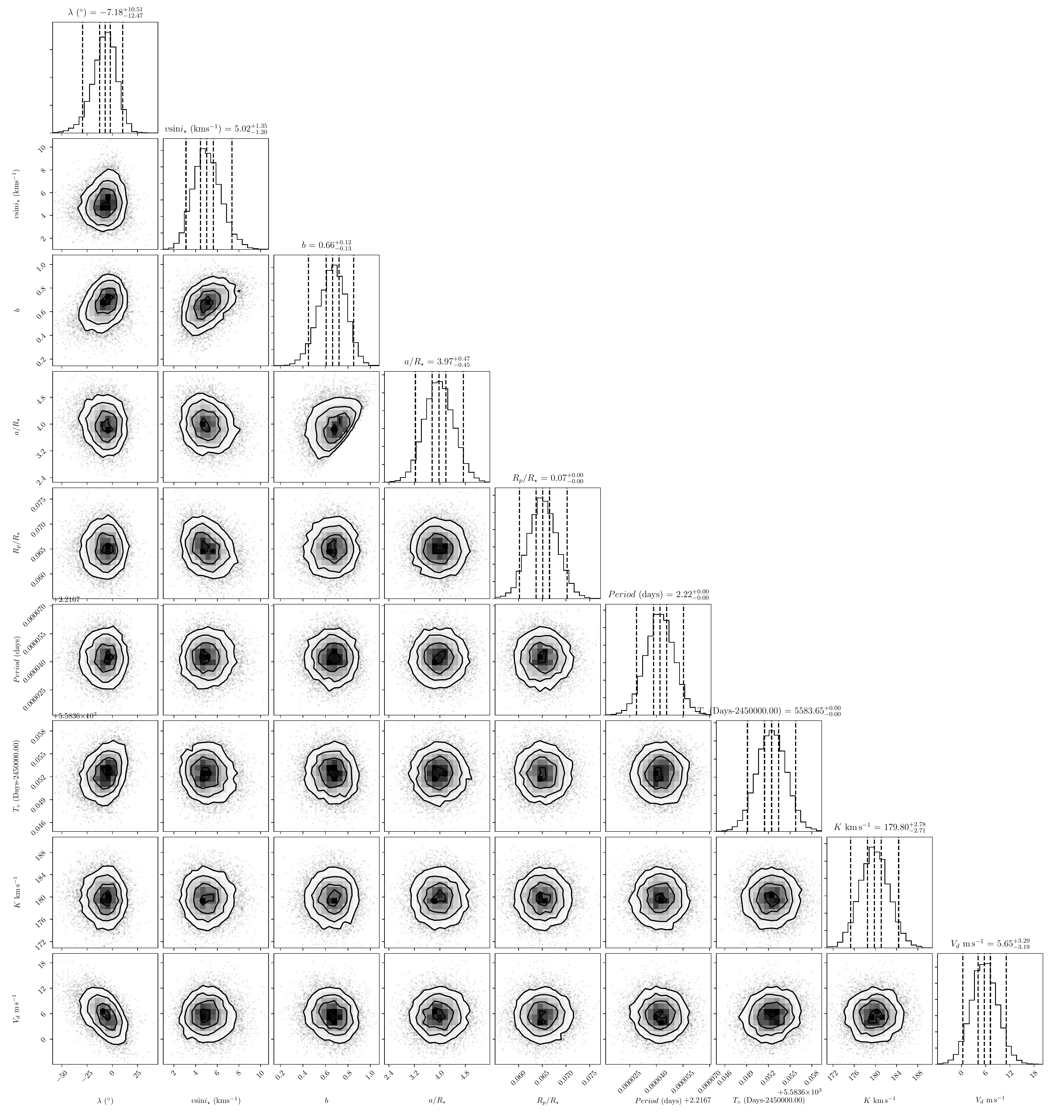}
	\caption{Corner distribution plot showing the potential correlations between the most relevant jump parameters used in the MCMC simulation of WASP-72. The distributions are mostly Gaussian indicating that only weak correlations exist between system parameters.}
	\label{fig:wasp-72_cor}
\end{figure}

Figure~\ref{fig:wasp-72_distro} shows the marginalized posterior probability distributions of $\lambda$ and $v\sin i_{\star}$ from the MCMC, which appears to adhere to a normal distribution. The $1\sigma$, $2\sigma$, and $3\sigma$ confidence contours are also plotted, along with normalized density functions marginalized over $\lambda$ and $v\sin i_{\star}$ with fitted Gaussians. Figure~\ref{fig:wasp-72_cor} is a corner distribution plot showing the correlations between all the modeled system parameters. No strong correlations are apparent in Figure~\ref{fig:wasp-72_cor}.

\begin{table*}
\centering
\begin{adjustbox}{max width=\textwidth}
\begin{threeparttable}[b]
\caption{System Parameters, Priors, and Results for WASP-72}
\centering
\begin{tabular}{l c c c c c}
\hline\hline \\ [-2.0ex]
Input Model Parameters & Prior & Prior Type & Results (normal $v\sin i_{\star}$ prior) & \textbf{Preferred Solution} (weak $3\sigma$ $v\sin i_{\star}$ prior) & Results (uniform $v\sin i_{\star}$ prior) \\ [0.5ex]
\hline \\ [-2.0ex]
Mid-transit epoch (2450000-HJD), $T_{0}$ & $5583.6529 \pm 0.0021$\tnote{a} & Gaussian & $5583.6524 \pm 0.0020$ & $5583.6524 \pm 0.0020$ & $5583.6524 \pm 0.0020$ \\

Orbital period (days), $P$ & $2.2167421 \pm 0.0000081$\tnote{a} & Gaussian & $2.2167420 \pm 0.0000080$ & $2.2167420 \pm 0.0000080$ & $2.2167420 \pm 0.0000080$ \\

Impact parameter, $b$ & $0.59^{+0.10}_{-0.18}$\tnote{a,b} & Gaussian & $0.69 \pm 0.11$ & $0.66 \pm 0.12$ & $0.65 \pm 0.13$ \\

Semi-major axis to star radius ratio, $a/R_{\star}$ & $4.02 \pm 0.49$\tnote{a} & Gaussian & $3.94 \pm 0.45$ & $3.98 \pm 0.46$ & $3.99 \pm 0.46$ \\

Planet-to-star radius ratio, $R_{P}/R_{\star}$ & $0.0656^{+0.0021}_{-0.0019}$\tnote{a,b} & Gaussian & $0.0647 \pm 0.0030$ & $0.0651 \pm 0.0031$ & $0.0653 \pm 0.0031$ \\

Orbital eccentricity, $e$ & $0$\tnote{c} & Fixed & -- & -- & -- \\

Argument of periastron, $\omega$ & --\tnote{c} & Fixed & -- & -- & -- \\

Stellar velocity semi-amplitude, $K$ & $181.0 \pm 4.2$\,\mos\tnote{a} & Gaussian & $179.8 \pm 2.7$\,\mos & $179.8 \pm 2.7$\,\mos  & $179.9 \pm 2.7$\,\mos\\

Stellar micro-turbulence, $\xi_{t}$ & N/A & Fixed & -- & -- & --  \\

Stellar macro-turbulence, $v_\mathrm{mac}$ & $4.0 \pm 0.3$\,\kms\tnote{a} & Gaussian & $4.0 \pm 0.3$\,\kms & $4.0 \pm 0.3$\,\kms & $4.0 \pm 0.3$\,\kms \\

Stellar limb-darkening coefficient, $q_{1}$ & $0.3990 \pm 0.0244$\tnote{d} & Gaussian & $0.3992 \pm 0.0244$ & $0.3993 \pm 0.0243$ & $0.3992 \pm 0.0243$ \\

Stellar limb-darkening coefficient, $q_{2}$ & $0.2679 \pm 0.0073$\tnote{d} & Gaussian & $0.2679 \pm 0.0073$ & $0.2679 \pm 0.0073$ & $0.2679 \pm 0.0073$ \\

RV data set offset\tnote{e}, $V_{d}$ & $[-50$ \textendash\, $50]$\,\mos & Uniform & $5.8 \pm 3.4$\,\mos & $5.7 \pm 3.3$\,\mos & $5.6 \pm 3.2$\,\mos \\

Projected obliquity angle, $\lambda$ & $[-60^{\circ}$ \textendash\, $60^{\circ}]$ & Uniform & ${-6^{\circ}}^{+10^{\circ}}_{-12^{\circ}}$ & ${-7^{\circ}}^{+11^{\circ}}_{-12^{\circ}}$ & ${-7^{\circ}}^{+11^{\circ}}_{-13^{\circ}}$ \\
 
Projected stellar rotation velocity, ${v\sin i_{\star}}$ & $6.00 \pm 0.70$\,\kms\tnote{a,f} & Gaussian & $5.8 \pm 0.7$\,\kms & $5.0^{+1.4}_{-1.2}$\,\kms & $4.7^{+1.7}_{-1.3}$\,\kms \\ [0.5ex]
\hline\hline \\ [-2.0ex]

Previously Derived Parameters (for informative purposes) & Value & -- & -- & -- & -- \\ [0.5ex]
\hline \\ [-2.0ex]
Orbital inclination, $I$ & ${81.6^{\circ}}^{+3.2^{\circ}}_{-2.6^{\circ}}$ & -- & -- & -- & -- \\
Stellar mass, $M_{\star}$ & $1.386 \pm 0.055$\,$M_{\odot}$ & -- & -- & -- & -- \\
Stellar radius, $R_{\star}$ & $1.98 \pm 0.24$\,$R_{\odot}$ & -- & -- & -- & --  \\
Planet mass, $M_{P}$ & $1.5461^{+0.059}_{-0.056}$\,$M_{J}$ & -- & -- & -- & --  \\
Planet radius, $R_{P}$ & $1.27 \pm 0.20$\,$R_{J}$ & -- & -- & -- & -- \\
\\ [0.5ex]
\hline 
\end{tabular}%
\vspace{1mm}
\label{table:WASP-72_Parameters}
\begin{tablenotes}
\item [a] \textit{Prior values given in \citet{2013A&A...552A..82G}.}
\item [b] \textit{In cases where the prior uncertainty is asymmetric, for simplicity, we use a symmetric Gaussian prior with the prior width set to the larger uncertainty value in the MCMC.}
\item [c] \textit{Fixed eccentricity to 0 as given by the preferred solution in \citet{2013A&A...552A..82G}.}
\item [d] \textit{Limb darkening coefficients interpolated from the look-up tables in \cite{2011A&A...529A..75C}.}
\item [e] \textit{RV offset between the \citet{2013A&A...552A..82G} and AAT data sets.}
\item [f] \textit{The uniform prior used for $v\sin i_{\star}$ is $U[1.0 - 12.0]$\,\kms\.}
\end{tablenotes}
\end{threeparttable}
\end{adjustbox}
\end{table*}

\subsection{WASP-100 Results} \label{sec:wasp-100_results}

\begin{figure}
	\centering
	\includegraphics[width=1.0\linewidth]{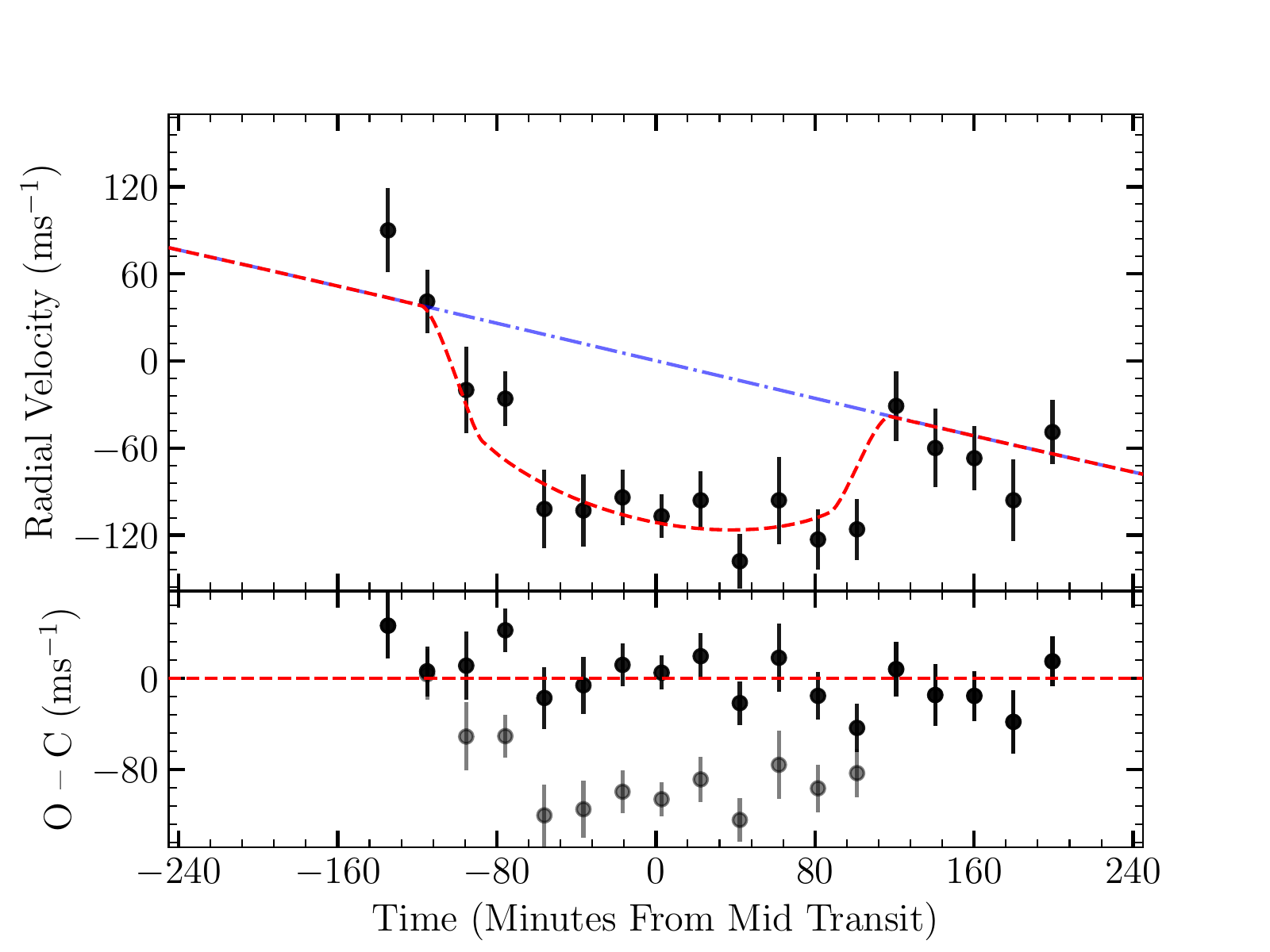}
	\caption{Spectroscopic radial velocities of the WASP-100 transit. Velocities from just before, during, and after the transit are plotted as a function of time (minutes from mid-transit at 2457298.114502\,HJD) along with the best fitting Rossiter-McLaughlin model (using the weak $3\sigma$ ${v\sin i_{\star}}$ prior, our preferred solution), Doppler model with no Rossister-McLaughlin effect, and corresponding residuals. The filled black circles with red error bars are radial velocities obtained in this work on 2015 October 2, the black circles in the residuals plot are from the best fit Rossiter-McLaughlin model, and the gray circles are the residuals from the Doppler model with no Rossister-McLaughlin effect. The velocity offset for the data set presented here was determined from the \citet{2014MNRAS.440.1982H} out-of-transit radial velocities.}
	\label{fig:WASP-100_RM}
\end{figure}

Figure~\ref{fig:WASP-100_RM} shows the observed RVs covering the full length of the WASP-100b transit, the best-fit modeled Rossiter-McLaughlin velocity anomaly, and the Doppler solution assuming no Rossiter-McLaughlin effect. In stark contrast to the situation of WASP-72b, the velocity anomaly measured for WASP-100b (see Figure~\ref{fig:WASP-100_RM}) strongly implies that the planet's orbit is significantly tilted (or even nearly polar) with respect to its host star's spin-axis. This is evident by the negative velocity anomaly observed over the entire duration of the transit, indicating that the planet transits across the red-shifted hemisphere from transit ingress to egress.

The best-fit projected spin-orbit angle for this system using the normal $v\sin i_{\star}$ prior of $v\sin i_{\star}=12.8 \pm 0.8$\,\kms\ is $\lambda = {79^{\circ}}^{+19^{\circ}}_{-10^{\circ}}$. Our preferred solution for $\lambda$ using the $3\sigma$ (weak) $v\sin i_{\star}$ prior of $v\sin i_{\star}=12.8 \pm 2.4$\,\kms\ results in $\lambda = {79^{\circ}}^{+19^{\circ}}_{-10^{\circ}}$. We also determined a solution for $\lambda$ using a uniform prior on $v\sin i_{\star}$, resulting in $\lambda = {80^{\circ}}^{+19^{\circ}}_{-11^{\circ}}$. The type of prior used for $v\sin i_{\star}$ has little influence on the $\lambda$ solution, again likely due to the high impact parameter of the transit of $b=0.64^{+0.08}_{-0.16}$. The solutions for the stellar rotation of WASP-100 are $v\sin i_{\star} = 12.8 \pm 0.8$\,\kms\, $v\sin i_{\star} = 12.8^{+2.3}_{-2.2}$\,\kms\, and $v\sin i_{\star} = 15.4^{+7.7}_{-5.6}$\,\kms\ for the normal, weak, and uniform $v\sin i_{\star}$ prior, respectively.

As an extra check to confirm the obvious Doppler anomaly signal in our time-series radial velocities, we have also calculated the BIC for WASP-100 and compared the BIC between the best fit (preferred solution) Rossiter-McLaughlin effect model and the Doppler model with no Rossiter-McLaughlin effect, finding $\Delta \mathrm{BIC}=213$. This provides decisive evidence in favor of the Rossiter-McLaughlin model.

\begin{figure}
	\centering
	\includegraphics[width=1.0\linewidth]{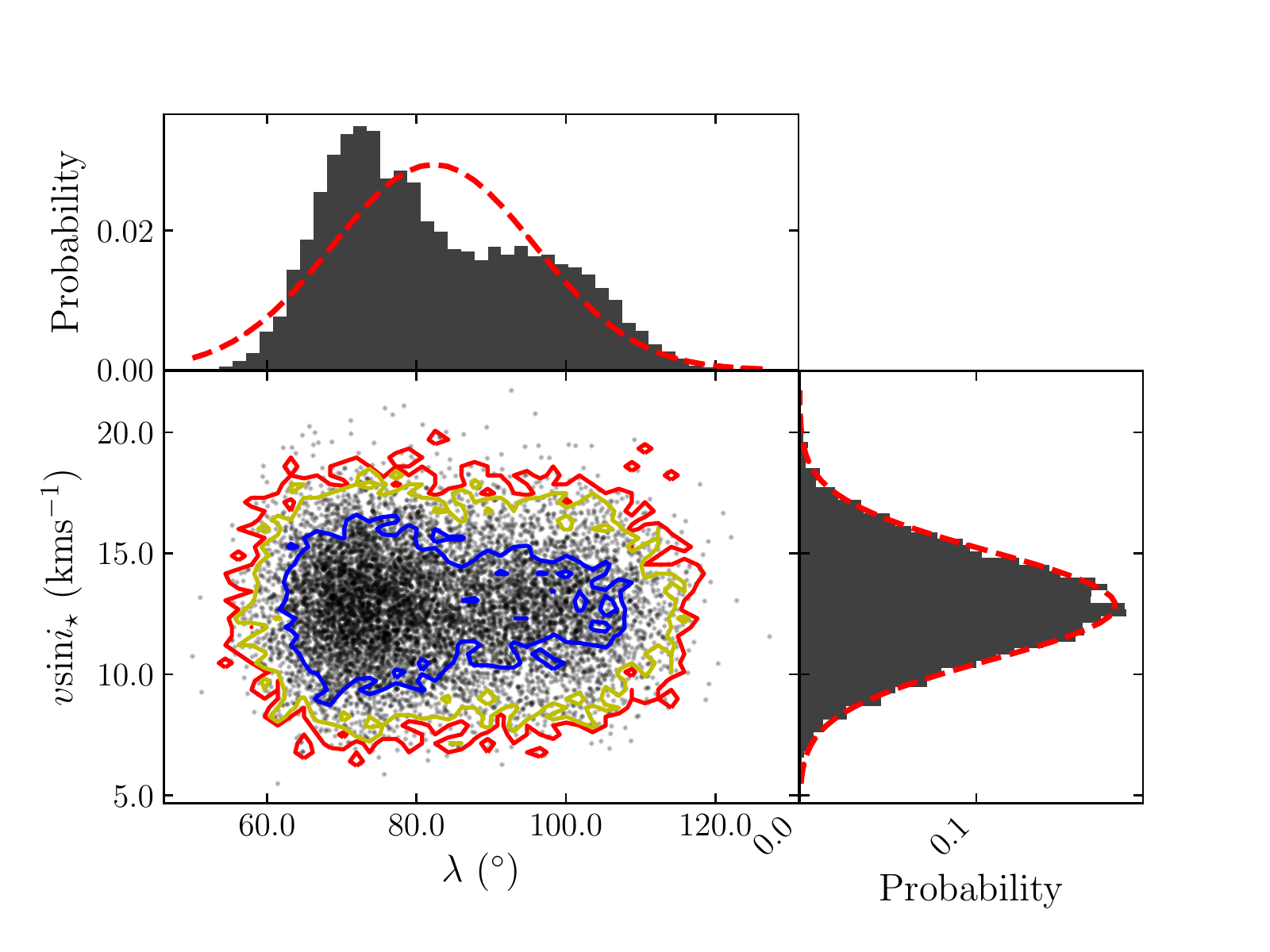}
	\caption{Posterior probability distribution of $\lambda$ and $v\sin i_{\star}$ from the MCMC simulation of WASP-100. The contours show the 1, 2, and 3 $\sigma$ confidence regions (in blue, yellow, and red, respectively). We have marginalized over $\lambda$ and $v\sin i_{\star}$ and have fit them with Gaussians (in red). This plot indicates that the distribution is somewhat non-Gaussian and suggest that there is a double-valued degenerate solution for $\lambda$.}
	\label{fig:wasp-100_distro}
\end{figure}

\begin{figure}
	\centering
	\includegraphics[width=1.0\linewidth]{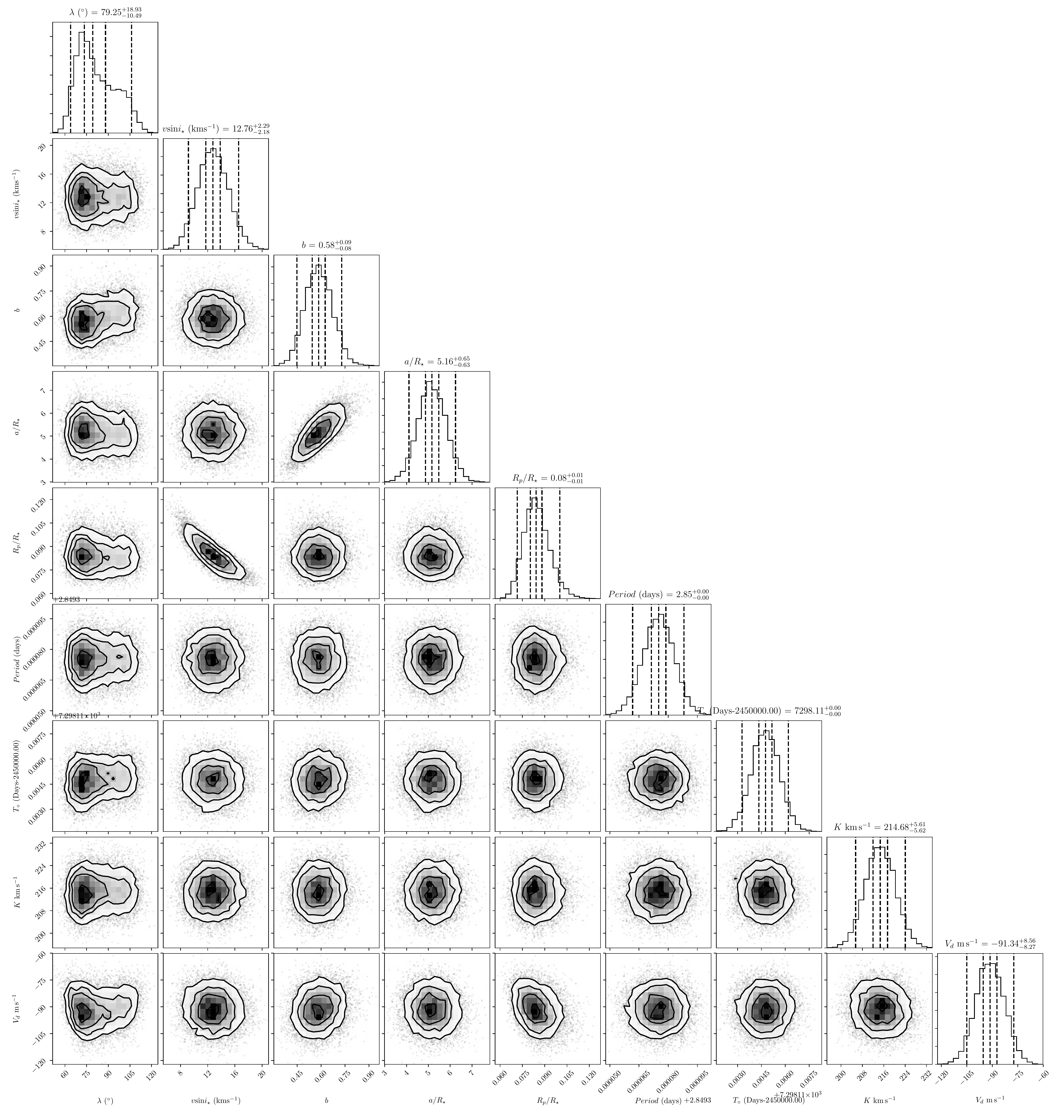}
	\caption{Corner distribution plot showing the potential correlations between the most relevant jump parameters used in the MCMC simulation of WASP-100. Some of the parameters do appear to be correlated with each other.}
	\label{fig:wasp-100_cor}
\end{figure}

We have plotted the posterior probability distributions from the MCMC fitting routine, marginalized over $\lambda$ and $v\sin i_{\star}$, in Figure~\ref{fig:wasp-100_distro}. The $1\sigma$, $2\sigma$, and $3\sigma$ confidence contours are also plotted, along with normalized density functions marginalized over $\lambda$ and $v\sin i_{\star}$ with fitted Gaussians. Figure~\ref{fig:wasp-100_distro} reveals that the distribution is somewhat non-Gaussian, elongated along the $\lambda$ axis with two possible peaks (the highest peaks near $\lambda = 75^{\circ}$ and the second peak near $\lambda = 100^{\circ}$), suggesting a double-valued degenerate solution. The cause of the double-valued degenerate solution is not known but might be from correlations between other system parameters, as evident between $R_{P}/R_{\star}$ and $v\sin i_{\star}$ and between $a//R_{\star}$ and $b$. This is shown in the series of correlation plots in Figure~\ref{fig:wasp-100_cor}.

\begin{table*}
\centering
\begin{adjustbox}{max width=\textwidth}
\begin{threeparttable}[b]
\caption{System Parameters, Priors, and Results for WASP-100}
\centering
\begin{tabular}{l c c c c c}
\hline\hline \\ [-2.0ex]
Input Model Parameters & Prior & Prior Type & Results (normal $v\sin i_{\star}$ prior) & \textbf{Preferred Solution} (weak $3\sigma$ $v\sin i_{\star}$ prior) & Results (uniform $v\sin i_{\star}$ prior) \\ [0.5ex]
\hline \\ [-2.0ex]
Mid-transit epoch (2450000-HJD), $T_{0}$ & $7298.1145 \pm 0.0009$\tnote{a} & Gaussian & $7298.1148 \pm 0.0009$ & $7298.1148 \pm 0.0009$ & $7298.1148 \pm 0.0009$ \\

Orbital period (days), $P$ & $2.849375 \pm 0.000008$\tnote{a} & Gaussian & $2.849375 \pm 0.000008$ & $2.849375 \pm 0.000008$ & $2.849375 \pm 0.000008$ \\

Impact parameter, $b$ & $0.64^{+0.08}_{-0.16}$\tnote{a,b} & Gaussian & $0.59 \pm 0.09$ & $0.59 \pm 0.09$ & $0.58 \pm 0.09$ \\

Semi-major axis to star radius ratio, $a/R_{\star}$ & $4.93 \pm 0.75$\tnote{a} & Gaussian & $5.18 \pm 0.66$ & $5.17 \pm 0.66$ & $5.15 \pm 0.66$ \\

Planet-to-star radius ratio, $R_{P}/R_{\star}$ & $0.0868 \pm 0.0224$\tnote{a} & Gaussian & $0.0841 \pm 0.0052$ & $0.0848 \pm 0.0088$ & $0.0789 \pm 0.0158$ \\

Orbital eccentricity, $e$ & $0$\tnote{c} & Fixed & -- & -- & -- \\

Argument of periastron, $\omega$ & --\tnote{c} & Fixed & -- & -- & -- \\

Stellar velocity semi-amplitude, $K$ & $213 \pm 8$\,\mos\tnote{a} & Gaussian & $215 \pm 6$\,\mos & $215 \pm 6$\,\mos  & $215 \pm 6$\,\mos\\

Stellar micro-turbulence, $\xi_{t}$ & N/A & Fixed & -- & -- & --  \\

Stellar macro-turbulence, $v_\mathrm{mac}$ & $[0.0$ \textendash\, $10.0]$\,\kms\tnote{d} & Uniform & $5.0 \pm 1.0$\,\kms & $5.0 \pm 1.0$\,\kms & $5.0 \pm 1.0$\,\kms \\

Stellar limb-darkening coefficient, $q_{1}$ & $0.2585 \pm 0.0064$\tnote{e} & Gaussian & $0.2585 \pm 0.0064$ & $0.2585 \pm 0.0063$ & $0.2585 \pm 0.0063$ \\

Stellar limb-darkening coefficient, $q_{2}$ & $0.3236 \pm 0.0066$\tnote{e} & Gaussian & $0.3236 \pm 0.0066$ & $0.3236 \pm 0.0066$ & $0.3235 \pm 0.0066$ \\

RV data set offset\tnote{f}, $V_{d}$ & $[-250$ \textendash\, $50]$\,\mos & Uniform & $-91 \pm 8$\,\mos & $-91 \pm 8$\,\mos & $-91 \pm 8$\,\mos \\

Projected obliquity angle, $\lambda$ & $[10^{\circ}$ \textendash\, $150^{\circ}]$ & Uniform & ${79^{\circ}}^{+19^{\circ}}_{-10^{\circ}}$ & ${79^{\circ}}^{+19^{\circ}}_{-10^{\circ}}$ & ${80^{\circ}}^{+19^{\circ}}_{-11^{\circ}}$ \\

Projected stellar rotation velocity, ${v\sin i_{\star}}$ & $12.8 \pm 0.8$\,\kms\tnote{a,g} & Gaussian & $12.8 \pm 0.8$\,\kms & $12.8^{+2.3}_{-2.2}$\,\kms & $15.4^{+7.7}_{-5.6}$\,\kms \\ [0.5ex]
\hline\hline \\ [-2.0ex]

Previously Derived Parameters (for informative purposes) & Value & -- & -- & -- & -- \\ [0.5ex]
\hline \\ [-2.0ex]
Orbital inclination, $I$ & ${82.6^{\circ}}^{+2.6^{\circ}}_{-1.7^{\circ}}$ & -- & -- & -- & -- \\
Stellar mass, $M_{\star}$ & $1.57 \pm 0.10$\,$M_{\odot}$ & -- & -- & -- & -- \\
Stellar radius, $R_{\star}$ & $2.0 \pm 0.3$\,$R_{\odot}$ & -- & -- & -- & --  \\
Planet mass, $M_{P}$ & $2.03 \pm 0.12$\,$M_{J}$ & -- & -- & -- & --  \\
Planet radius, $R_{P}$ & $1.69 \pm 0.29$\,$R_{J}$ & -- & -- & -- & -- \\
\\ [0.5ex]
\hline 
\end{tabular}%
\vspace{1mm}
\label{table:WASP-100_Parameters}
\begin{tablenotes}
\item [a] \textit{Prior values given in \citet{2014MNRAS.440.1982H}.}
\item [b] \textit{In cases where the prior uncertainty is asymmetric, for simplicity, we use a symmetric Gaussian prior with the prior width set to the larger uncertainty value in the MCMC.}
\item [c] \textit{Fixed eccentricity to 0 as given by the preferred solution in \citet{2014MNRAS.440.1982H}.}
\item [d] \textit{No prior value for the macro-turbulence parameter given in \citet{2014MNRAS.440.1982H}. We used a uniform prior on the given interval.}
\item [e] \textit{Limb darkening coefficients interpolated from the look-up tables in \cite{2011A&A...529A..75C}.}
\item [f] \textit{RV offset between the \citet{2014MNRAS.440.1982H} and AAT data sets.}
\item [g] \textit{The uniform prior used for $v\sin i_{\star}$ is $U[5.0 - 30.0]$\,\kms\.}
\end{tablenotes}
\end{threeparttable}
\end{adjustbox}
\end{table*}

\subsection{WASP-109 Results} \label{sec:wasp-109_results}

\begin{figure}
	\centering
	\includegraphics[width=1.0\linewidth]{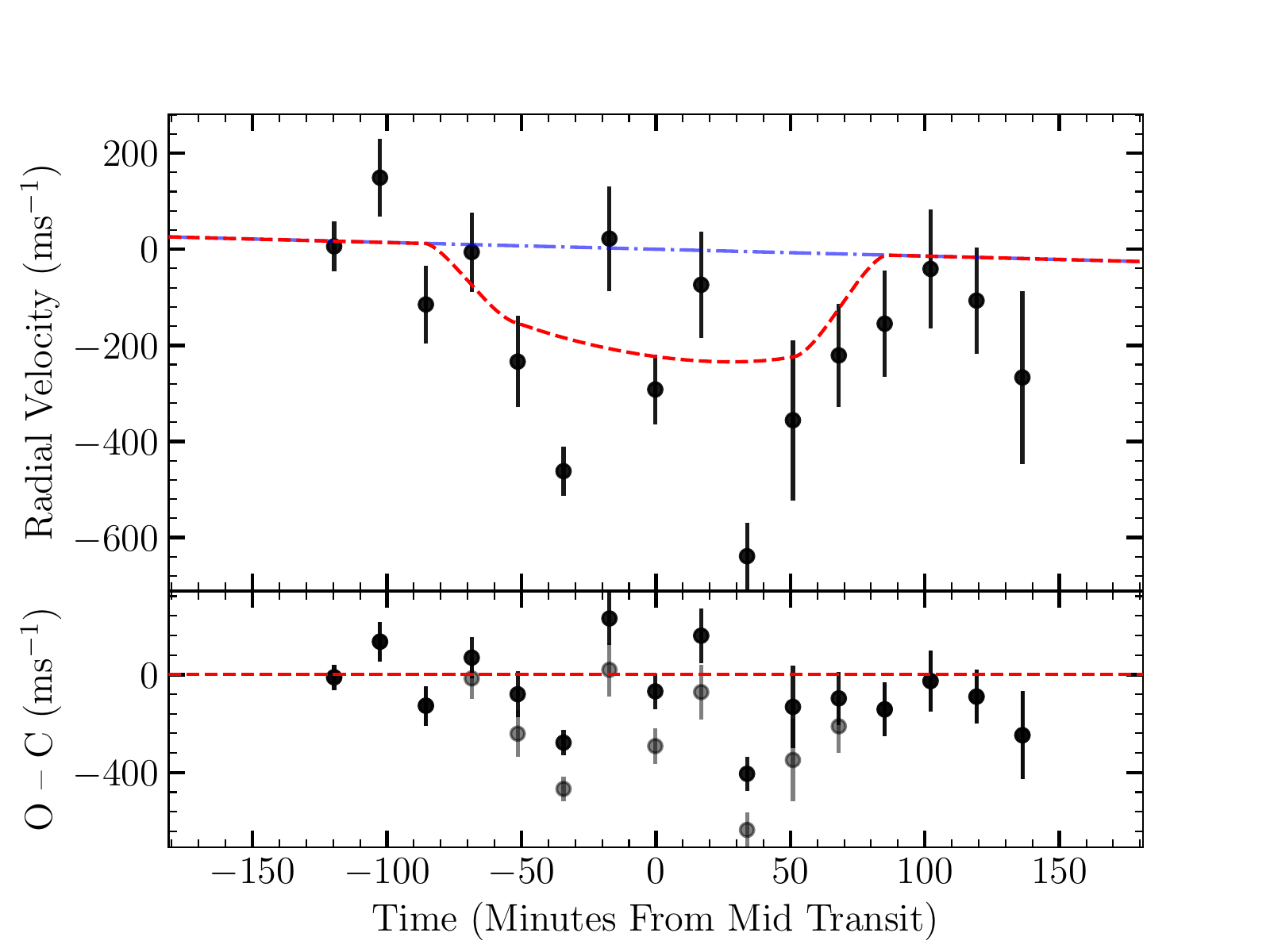}
	\caption{Spectroscopic radial velocities of the WASP-109 transit. Velocities from just before, during, and after the transit are plotted as a function of time (minutes from mid-transit at 2457151.1201754\,HJD) along with the best fitting Rossiter-McLaughlin model (using the weak $3\sigma$ ${v\sin i_{\star}}$ prior, our preferred solution), Doppler model with no Rossister-McLaughlin effect, and corresponding residuals. The filled black circles with red error bars are radial velocities obtained in this work on 2015 May 8, the black circles in the residuals plot are from the best fit Rossiter-McLaughlin model, and the gray circles are the residuals from the Doppler model with no Rossister-McLaughlin effect. The velocity offset for the data set presented here was determined from the \citet{2014arXiv1410.3449A} out-of-transit radial velocities.}
	\label{fig:WASP-109_RM}
\end{figure}

Similar to the case of WASP-100b, WASP-109b also appears to exhibit a highly inclined orbit with respect to its host star's projected spin-axis. As shown in Figure~\ref{fig:WASP-109_RM}, the Rossiter-McLaughlin effect appears as a negative velocity anomaly during the transit. However, some caution is needed with interpreting these results as there is an unusual amount of radial velocity scatter in the residuals to the Rossiter-McLaughlin best-fit model (as shown on the bottom of Figure~\ref{fig:WASP-109_RM}). We acknowledge that our time-series radial velocities of WASP-109 could contain correlated ('red') noise and/or systematics that have not been taken into account since more radial velocities lie below the best-fit line than above it. We would have also benefited from additional out-of-transit radial velocity measurements, additional in-transit radial velocities, and Doppler tomography analysis \citep[e.g., see][]{2017AJ....154..137J} of this system.

Despite the potential unaccounted for systematics in our radial velocity measurements, we determined the best-fit projected spin-orbit angle as $\lambda = 99^{\circ} \pm -10^{\circ}$ using the normal $v\sin i_{\star}$ prior of $v\sin i_{\star}=15.4 \pm 1.0$\,\kms\. Our preferred solution for $\lambda$ using the $3\sigma$ (weak) $v\sin i_{\star}$ prior of $v\sin i_{\star}=15.4 \pm 3.0$\,\kms\ results in $\lambda = {99^{\circ}}^{+10^{\circ}}_{-9^{\circ}}$. We also determined a solution for $\lambda$ using a uniform prior on $v\sin i_{\star}$, resulting in $\lambda = 100^{\circ} \pm 8^{\circ}$. The solution for $\lambda$ appears to be independent of the $v\sin i_{\star}$ prior we used due to the high impact parameter of the transit of $b=0.737 \pm 0.011$. This is likely the reason for our precise determination of $\lambda$ even with the high level of radial velocity scatter in the residuals. Additionally, we have also calculated the BIC for WASP-109 and compared the BIC between our best fit (preferred solution) Rossiter-McLaughlin effect model and the Doppler model with no Rossiter-McLaughlin effect, finding $\Delta \mathrm{BIC}=116$ in favor of the Rossiter-McLaughlin model.

The solutions for the stellar rotation of WASP-109 are $v\sin i_{\star} = 16.0^{+1.0}_{-0.9}$\,\kms\, $v\sin i_{\star} = 18.9^{+2.4}_{-2.3}$\,\kms\, and $v\sin i_{\star} = 29.6^{+5.7}_{-5.9}$\,\kms\ for the normal, weak, and uniform $v\sin i_{\star}$ prior, respectively. Using a uniform prior on $v\sin i_{\star}$ results in unreasonably large value for $v\sin i_{\star}$ ($\sim 2.4\sigma$ from the reported value of $v\sin i_{\star}=15.4 \pm 1.0$\,\kms\ in \citealt{2014arXiv1410.3449A}). While the uniform prior on $v\sin i_{\star}$ does result in a better fit to the data (BIC of 63 compared to a BIC of 89 using the weak $v\sin i_{\star}$), in general, Rossiter-McLaughlin observations only provide weak constraints on the stellar rotational velocity. External data can provide much more leverage for measuring $v\sin i_{\star}$, such as from using high S/N, high-resolution out-of-transit spectroscopy to determine $v\sin i_{\star}$. Therefore, our preferred solution for all three systems makes use of the prior information on $v\sin i_{\star}$ by placing a $3\sigma$ prior on this parameter though we have also included the solutions using a normal and uniform prior on $v\sin i_{\star}$.

\begin{figure}
	\centering
	\includegraphics[width=1.0\linewidth]{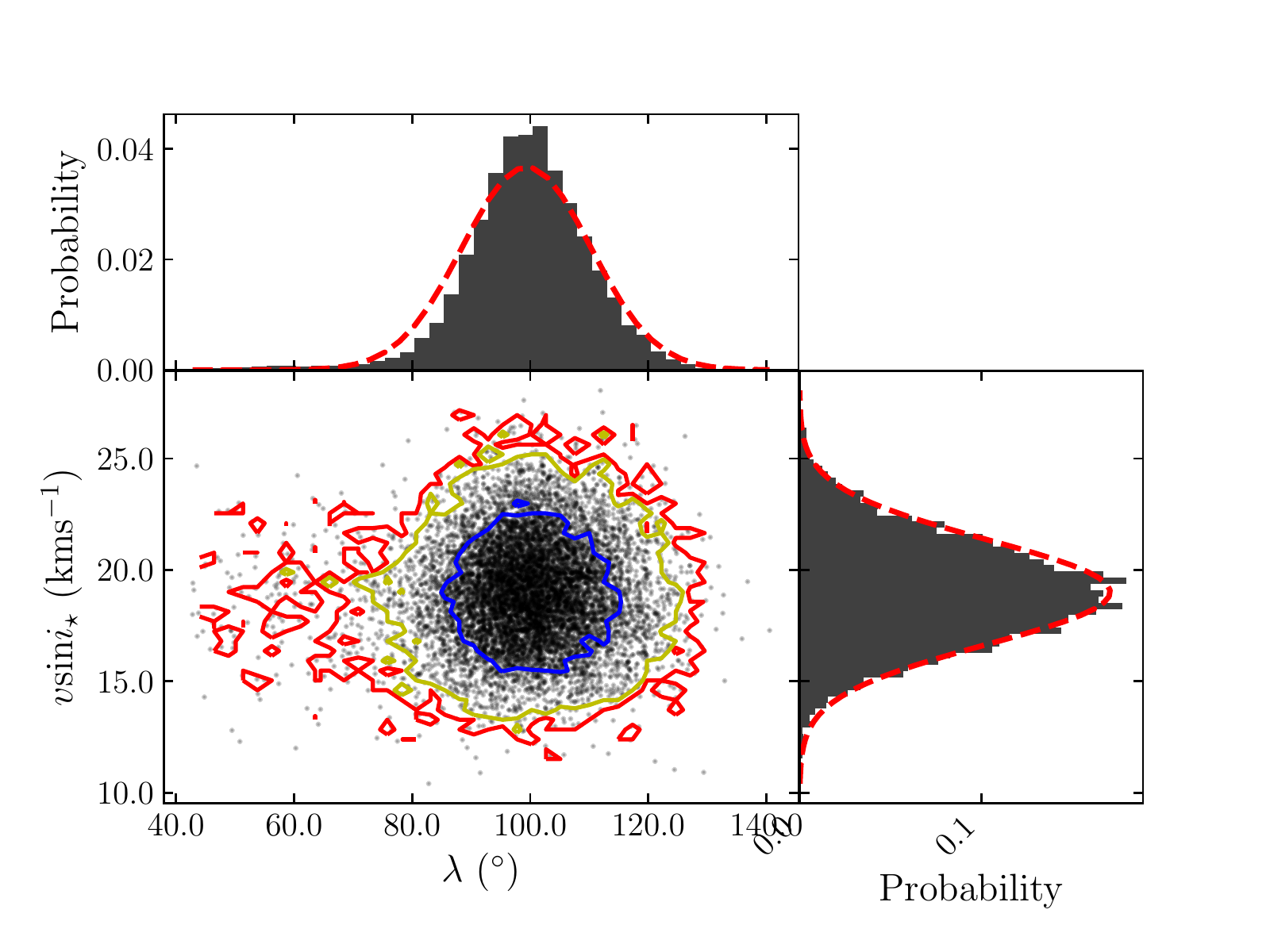}
	\caption{Posterior probability distribution of $\lambda$ and $v\sin i_{\star}$ from the MCMC simulation of WASP-109. The contours show the 1, 2, and 3 $\sigma$ confidence regions (in blue, yellow, and red, respectively). We have marginalized over $\lambda$ and $v\sin i_{\star}$ and have fit them with Gaussians (in red). This plot indicates that the distribution is mostly Gaussian suggesting only a weak correlation between $\lambda$ and $v\sin i_{\star}$.}
	\label{fig:wasp-109_distro}
\end{figure}

\begin{figure}
	\centering
	\includegraphics[width=1.0\linewidth]{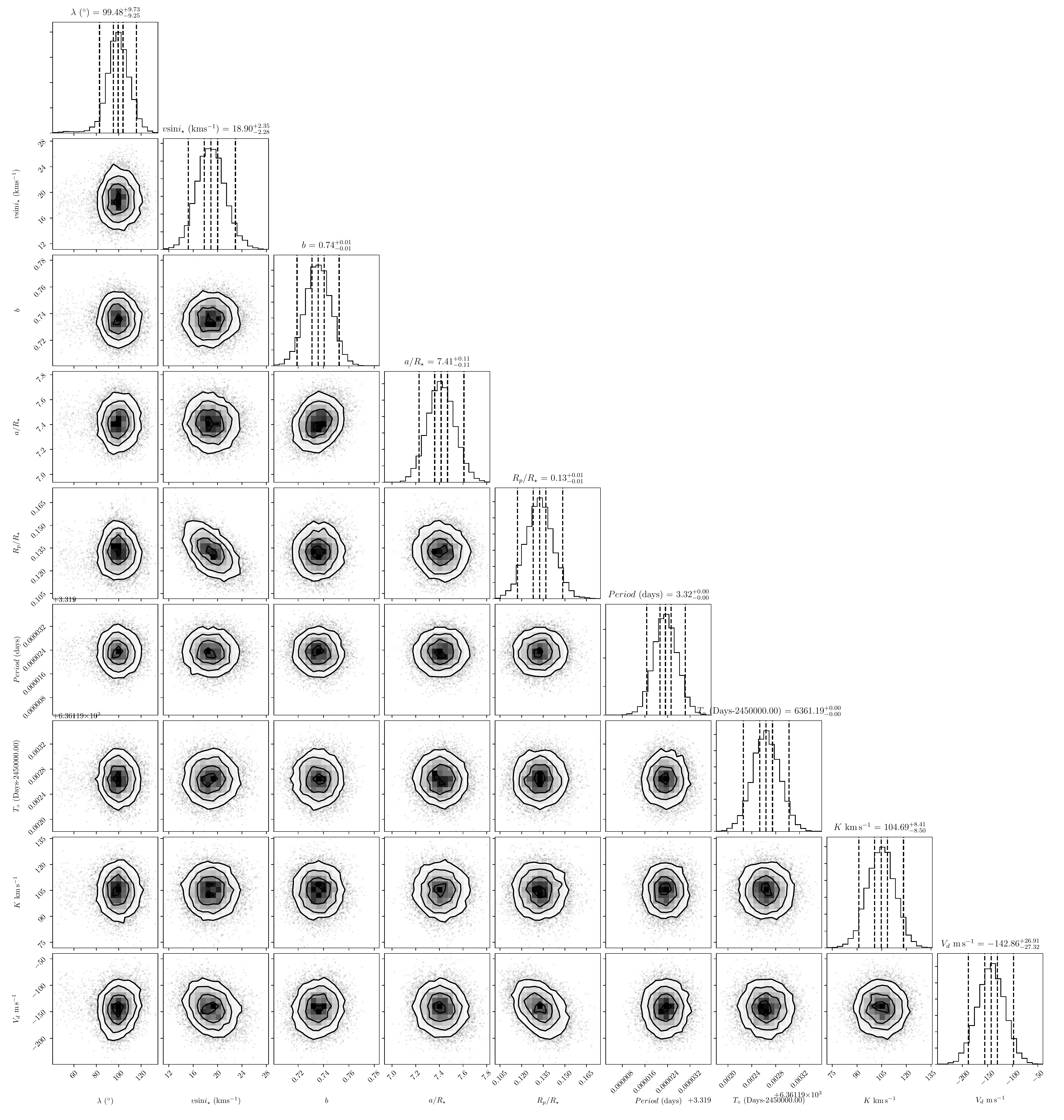}
	\caption{Corner distribution plot showing the potential correlations between the most relevant jump parameters used in the MCMC simulation of WASP-109. Some of the parameters do appear to be somewhat correlated with each other.}
	\label{fig:wasp-109_cor}
\end{figure}

The posterior probability distributions from the MCMC, marginalized over $\lambda$ and $v\sin i_{\star}$, are shown in Figure~\ref{fig:wasp-109_distro}, similar to Figures~\ref{fig:wasp-72_distro} and \ref{fig:wasp-100_distro}. The distribution is fairly Gaussian shaped with a trailing tail of lightly populated samples along lower $\lambda$ values. Figure~\ref{fig:wasp-109_cor} is a corner distribution plot showing the correlations between all the modeled system parameters. $R_{P}/R_{\star}$ and $v\sin i_{\star}$ appear to be weakly correlated and might explain the trailing tail observed in Figure~\ref{fig:wasp-109_distro}.

\begin{table*}
\centering
\begin{adjustbox}{max width=\textwidth}
\begin{threeparttable}[b]
\caption{System Parameters, Priors, and Results for WASP-109}
\centering
\begin{tabular}{l c c c c c}
\hline\hline \\ [-2.0ex]
Input Model Parameters & Prior & Prior Type & Results (normal $v\sin i_{\star}$ prior) & \textbf{Preferred Solution} (weak $3\sigma$ $v\sin i_{\star}$ prior) & Results (uniform $v\sin i_{\star}$ prior) \\ [0.5ex]
\hline \\ [-2.0ex]
Mid-transit epoch (2450000-HJD), $T_{0}$ & $6361.19263 \pm 0.00023$\tnote{a} & Gaussian & $6361.19263 \pm 0.00023$ & $6361.19263 \pm 0.00023$ & $6361.19263 \pm 0.00023$ \\

Orbital period (days), $P$ & $3.3190233 \pm 0.0000042$\tnote{a} & Gaussian & $3.3190233 \pm 0.0000040$ & $3.3190233 \pm 0.0000040$ & $3.3190233 \pm 0.0000040$ \\

Impact parameter, $b$ & $0.737 \pm 0.011$\tnote{a} & Gaussian & $0.735 \pm 0.010$ & $0.736 \pm 0.010$ & $0.738 \pm 0.010$ \\

Semi-major axis to star radius ratio, $a/R_{\star}$ & $7.40 \pm 0.13$\tnote{a} & Gaussian & $7.42 \pm 0.11$ & $7.42 \pm 0.12$ & $7.39 \pm 0.12$ \\

Planet-to-star radius ratio, $R_{P}/R_{\star}$ & $0.1101 \pm 0.0138$\tnote{a} & Gaussian & $0.1390 \pm 0.0091$ & $0.1327 \pm 0.0096$ & $0.1141 \pm 0.0109$ \\

Orbital eccentricity, $e$ & $0$\tnote{b} & Fixed & -- & -- & -- \\

Argument of periastron, $\omega$ & --\tnote{b} & Fixed & -- & -- & -- \\

Stellar velocity semi-amplitude, $K$ & $109 \pm 15$\,\mos\tnote{a} & Gaussian & $105 \pm 8$\,\mos & $105 \pm 8$\,\mos  & $104 \pm 8$\,\mos\\

Stellar micro-turbulence, $\xi_{t}$ & N/A & Fixed & -- & -- & --  \\

Stellar macro-turbulence, $v_\mathrm{mac}$ & $6.5 \pm 0.6$\,\kms\tnote{a} & Gaussian & $6.5 \pm 0.6$\,\kms & $6.5 \pm 0.6$\,\kms & $6.5 \pm 0.6$\,\kms \\

Stellar limb-darkening coefficient, $q_{1}$ & $0.3710 \pm 0.0186$\tnote{c} & Gaussian & $0.3710 \pm 0.0186$ & $0.3710 \pm 0.0186$ & $0.3710 \pm 0.0184$ \\

Stellar limb-darkening coefficient, $q_{2}$ & $0.2785 \pm 0.0041$\tnote{c} & Gaussian & $0.2786 \pm 0.0041$ & $0.2785 \pm 0.0041$ & $0.2785 \pm 0.0041$ \\

RV data set offset\tnote{d}, $V_{d}$ & $[-350$ \textendash\, $50]$\,\mos & Uniform & $-135 \pm 27$\,\mos & $-143 \pm 27$\,\mos & $-159 \pm 27$\,\mos \\

Projected obliquity angle, $\lambda$ & $[0^{\circ}$ \textendash\, $180^{\circ}]$ & Uniform & $99^{\circ} \pm 10^{\circ}$ & ${99^{\circ}}^{+10^{\circ}}_{-9^{\circ}}$ & $100 \pm 8^{\circ}$ \\

Projected stellar rotation velocity, ${v\sin i_{\star}}$ & $15.4 \pm 1.0$\,\kms\tnote{a,g} & Gaussian & $16.0^{+1.0}_{-0.9}$\,\kms & $18.9^{+2.4}_{-2.3}$\,\kms & $29.6^{+5.7}_{-5.9}$\,\kms \\ [0.5ex]
\hline\hline \\ [-2.0ex]

Previously Derived Parameters (for informative purposes) & Value & -- & -- & -- & -- \\ [0.5ex]
\hline \\ [-2.0ex]
Orbital inclination, $I$ & ${84.28^{\circ} \pm 0.19^{\circ}}$ & -- & -- & -- & -- \\
Stellar mass, $M_{\star}$ & $1.203 \pm 0.090$\,$M_{\odot}$ & -- & -- & -- & -- \\
Stellar radius, $R_{\star}$ & $1.346 \pm 0.044$\,$R_{\odot}$ & -- & -- & -- & --  \\
Planet mass, $M_{P}$ & $0.91 \pm 0.13$\,$M_{J}$ & -- & -- & -- & --  \\
Planet radius, $R_{P}$ & $1.443 \pm 0.053$\,$R_{J}$ & -- & -- & -- & -- \\
\\ [0.5ex]
\hline 
\end{tabular}%
\vspace{1mm}
\label{table:WASP-109_Parameters}
\begin{tablenotes}
\item [a] \textit{Prior values given in \citet{2014arXiv1410.3449A}.}
\item [b] \textit{Fixed eccentricity to 0 as given by the preferred solution in \citet{2014arXiv1410.3449A}.}
\item [c] \textit{Limb darkening coefficients interpolated from the look-up tables in \cite{2011A&A...529A..75C}.}
\item [d] \textit{RV offset between the \citet{2014arXiv1410.3449A} and AAT data sets.}
\item [g] \textit{The uniform prior used for $v\sin i_{\star}$ is $U[10.0 - 40.0]$\,\kms\.}
\end{tablenotes}
\end{threeparttable}
\end{adjustbox}
\end{table*}

\section{Discussion}\label{sec:Discussion_wasp-72-109}
 
\begin{figure*}
	\centering
	\includegraphics[width=1.0\linewidth]{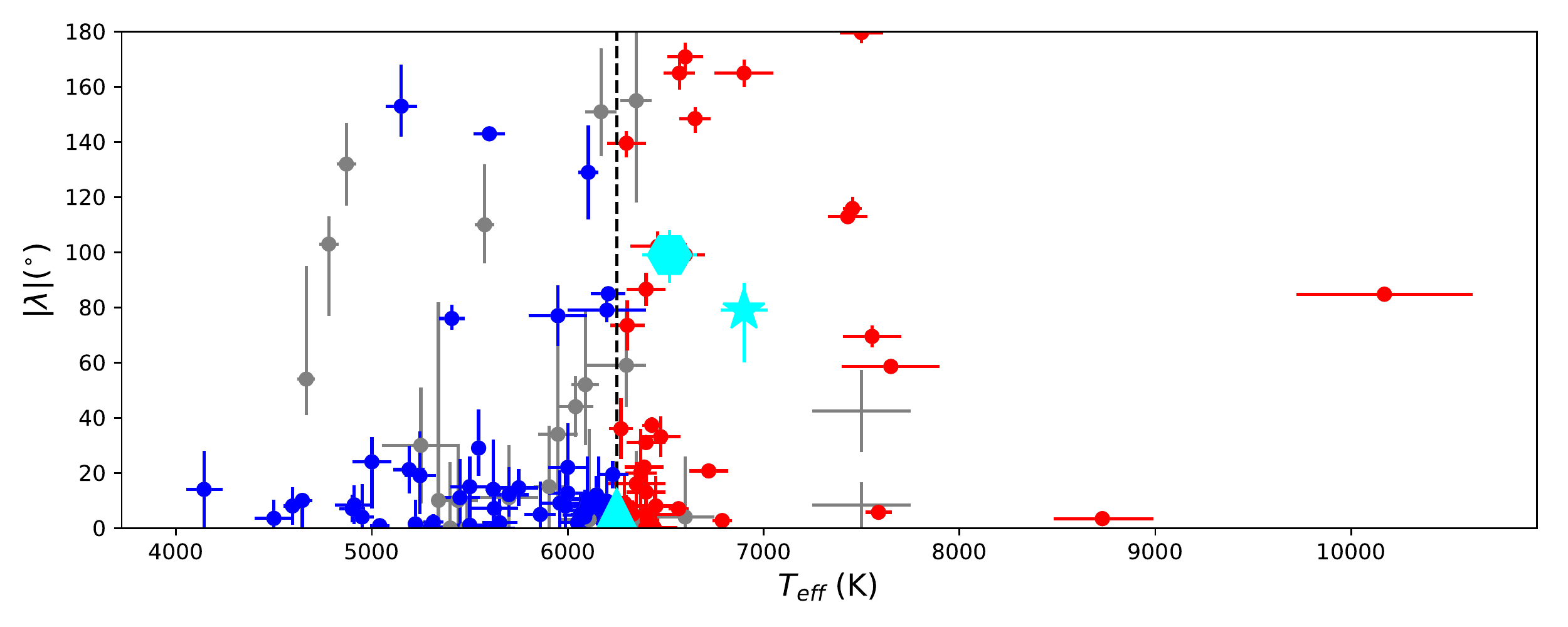}
	\caption{Distribution of spin-orbit misalignments $\lambda$ as a function of $T_{\mathrm{eff}}$ for hot Jupiters ($P<10$ days, $0.3 M_J<M_P<13 M_J$) from the literature. Planets orbiting stars below and above the Kraft break ($T_{\mathrm{eff}}=6250$ K, marked by the vertical dashed line) are shown with blue and red points, respectively, while gray points show planets with uncertainties of more than $20^{\circ}$ on the published values of $\lambda$. Our measurements for WASP-72b, WASP-100b, and WASP-109b are denoted by the cyan triangle, star, and hexagon, respectively. The literature sample was assembled using John Southworth's TEPCat Rossiter-McLaughlin Catalogue\protect{\footnote{http://www.astro.keele.ac.uk/jkt/tepcat/}}.}
	\label{fig:spinorbit}
\end{figure*}

Our measurements of the spin-orbit misalignments for WASP-72b, -100b, and -109b add to the several dozen such measurements now available in the literature (shown in Fig.~\ref{fig:spinorbit}). The picture initially presented by \cite{Winn:2010} has largely stood the test of time: hot Jupiters orbiting stars below the Kraft break tend to have aligned orbits (with only a few exceptions, most of which are at large $a/R_{\star}$, where tidal damping is less effective), while those above the Kraft break have a wide distribution of misalignments. Our new measurements fit into this picture well. WASP-72, with $T_{\mathrm{eff}}=6250\pm120$ K, is located at the Kraft break, and its hot Jupiter has a well-aligned orbit ($\lambda={-7^{\circ}}^{+11^{\circ}}_{-12^{\circ}}$). WASP-100b and WASP-109b both orbit somewhat hotter stars ($T_{\mathrm{eff}}=6900\pm120$ and $6520\pm140$ K, respectively), and both have highly inclined, polar orbits ($\lambda={79^{\circ}}^{+19^{\circ}}_{-10^{\circ}}$ and $\lambda={99^{\circ}}^{+10^{\circ}}_{-9^{\circ}}$, respectively).

Each of the dynamical migration mechanisms mentioned in the introduction predict a different distribution of $\lambda$ for hot Jupiters, and so measuring this distribution will allow us to distinguish between different predicted misalignment mechanisms. An initial attempt at such an analysis was performed by \cite{MortonJohnson:2011}, but the sample at that time was insufficient to produce a robust result. Only by measuring additional spin-orbit alignments of stars above the Kraft break (as we have done for WASP-100 and WASP-109) can we produce an observed distribution of spin-orbit alignments which is likely to be reflective of the primordial distribution, as these planets should have experienced minimal tidal damping \citep[e.g.,][]{Dawson:2014}.

Planets with significant spin-orbit misalignments ($|\lambda|>40^{\circ}$) are particularly important as in the case of more aligned orbits it is difficult to distinguish between planets that were originally emplaced onto aligned orbits, and those that experienced tidal realignment \citep[e.g.,][]{CridaBatygin:2014}. WASP-100b and -109b add to this number, and thus will be valuable for analyses of the hot Jupiter population as a whole. There are now 40 hot Jupiters orbiting stars with $T_{\mathrm{eff}}>6250$ K at $1\sigma$ confidence and which have $\lambda$ measured to a precision of 20$^{\circ}$ or better, 16 of which are significantly misaligned. This is approaching the number of measurements that \cite{MortonJohnson:2011} found would be necessary in order to confidently distinguish between models of Kozai-Lidov versus planetary scattering for hot Jupiter migration. A reassessment of this issue in the near future would therefore be valuable; however, given the possibility that not all hot Jupiters are produced by the same migration mechanism, even more spin-orbit misalignment measurements will likely be needed before this issue can be fully settled. Such an analysis is beyond the scope of this work but we encourage this work in the near future.



\section{Conclusions}\label{sec:Conclusions_wasp-66-103}
We have determined the sky-projected spin-orbit angle of three transiting hot Jupiter systems from spectroscopic observations of the Rossiter-McLaughlin effect obtain on the Anglo-Australian Telescope using the CYCLOPS2 fiber-feed. These observations reveal that WASP-100b and WASP-109b are on highly misaligned, nearly polar orbits of $\lambda = {79^{\circ}}^{+19^{\circ}}_{-10^{\circ}}$ and $\lambda = {99^{\circ}}^{+10^{\circ}}_{-9^{\circ}}$, respectively. In contrast, WASP-72b appears to be on an orbit that is aligned with its host star's equator ($\lambda = {-7^{\circ}}^{+11^{\circ}}_{-12^{\circ}}$).

The spin-orbit angles of these systems follow the trend first presented by \cite{Winn:2010} -- stars hotter than $T_{\mathrm{eff}}\sim6250$\,K host the majority of hot Jupiters on misaligned orbits. This temperature boundary corresponds to the Kraft break, which separates stars with deep convective envelopes that can effectively tidally realign planetary orbits (those cooler than $T_{\mathrm{eff}}\sim6250$\,K) and stars that have thin convective envelopes. WASP-100b and WASP-109b orbit hosts above the Kraft break while WASP-72b orbits a host that has an effective temperature at the boundary.

We are now approaching the number of measurements that are necessary to distinguish between planetary migration model for hot Jupiters. A statistical analysis of the ensemble of hot Jupiter systems will be valuable in future studies, especially once TESS begins discovering hundreds of new planets orbiting bright stars \citep{2014SPIE.9143E..20R}.

\clearpage

\acknowledgments

We acknowledge the use of the SIMBAD database, operated at CDS, Strasbourg, France. This research has made use of NASA's Astrophysics Data System, the Ren\'{e} Heller's Holt-Rossiter--McLaughlin Encyclopaedia (\url{http://www.astro.physik.uni-goettingen.de/~rheller/}), John Southworth's TEPCat Rossiter-McLaughlin Catalogue (\url{http://www.astro.keele.ac.uk/jkt/tepcat/}), the Exoplanet Orbit Database and the Exoplanet Data Explorer at \url{exoplanets.org}, and the Extrasolar Planets Encyclopaedia at \url{http://exoplanet.eu}.
S.W. thanks the Heising-Simons Foundation for their generous support. We thank the referee for the insightful comments and suggestions for improving this manuscript.

\bibliography{WASP-72_WASP-100_WASP-109}

\clearpage

\end{document}

%% file: aas_macros.tex
\def\jnl@style{\it}
\def\aaref@jnl#1{{\jnl@style#1}}

\def\aaref@jnl#1{{\jnl@style#1}}

\def\aj{\aaref@jnl{AJ}}                   
\def\araa{\aaref@jnl{ARA\&A}}             
\def\apj{\aaref@jnl{ApJ}}                 
\def\apjl{\aaref@jnl{ApJ}}                
\def\apjs{\aaref@jnl{ApJS}}               
\def\ao{\aaref@jnl{Appl.~Opt.}}           
\def\apss{\aaref@jnl{Ap\&SS}}             
\def\aap{\aaref@jnl{A\&A}}                
\def\aapr{\aaref@jnl{A\&A~Rev.}}          
\def\aaps{\aaref@jnl{A\&AS}}              
\def\azh{\aaref@jnl{AZh}}                 
\def\baas{\aaref@jnl{BAAS}}               
\def\jrasc{\aaref@jnl{JRASC}}             
\def\memras{\aaref@jnl{MmRAS}}            
\def\mnras{\aaref@jnl{MNRAS}}             
\def\pra{\aaref@jnl{Phys.~Rev.~A}}        
\def\prb{\aaref@jnl{Phys.~Rev.~B}}        
\def\prc{\aaref@jnl{Phys.~Rev.~C}}        
\def\prd{\aaref@jnl{Phys.~Rev.~D}}        
\def\pre{\aaref@jnl{Phys.~Rev.~E}}        
\def\prl{\aaref@jnl{Phys.~Rev.~Lett.}}    
\def\pasp{\aaref@jnl{PASP}}               
\def\pasj{\aaref@jnl{PASJ}}               
\def\jsara{\aaref@jnl{JSARA}}             
\def\qjras{\aaref@jnl{QJRAS}}             
\def\skytel{\aaref@jnl{S\&T}}             
\def\solphys{\aaref@jnl{Sol.~Phys.}}      
\def\sovast{\aaref@jnl{Soviet~Ast.}}      
\def\ssr{\aaref@jnl{Space~Sci.~Rev.}}     
\def\zap{\aaref@jnl{ZAp}}                 
\def\nat{\aaref@jnl{Nature}}              
\def\iaucirc{\aaref@jnl{IAU~Circ.}}       
\def\aplett{\aaref@jnl{Astrophys.~Lett.}} 
\def\apspr{\aaref@jnl{Astrophys.~Space~Phys.~Res.}}
\def\bain{\aaref@jnl{Bull.~Astron.~Inst.~Netherlands}} 
\def\fcp{\aaref@jnl{Fund.~Cosmic~Phys.}}  
\def\gca{\aaref@jnl{Geochim.~Cosmochim.~Acta}}   
\def\grl{\aaref@jnl{Geophys.~Res.~Lett.}} 
\def\jcp{\aaref@jnl{J.~Chem.~Phys.}}      
\def\jgr{\aaref@jnl{J.~Geophys.~Res.}}    
\def\jqsrt{\aaref@jnl{J.~Quant.~Spec.~Radiat.~Transf.}}
\def\memsai{\aaref@jnl{Mem.~Soc.~Astron.~Italiana}}
\def\nphysa{\aaref@jnl{Nucl.~Phys.~A}}   
\def\physrep{\aaref@jnl{Phys.~Rep.}}   
\def\physscr{\aaref@jnl{Phys.~Scr}}   
\def\planss{\aaref@jnl{Planet.~Space~Sci.}}   
\def\procspie{\aaref@jnl{Proc.~SPIE}}   
\def\nar{\aaref@jnl{NewAR}}   
\def\icarus{\aaref@jnl{Icarus}}